\documentclass{ieeetj}

\usepackage{xcolor,soul,framed} %

\usepackage[noadjust,sort]{cite}
\usepackage{url}

\usepackage{tabularx}

\usepackage{comment}

\usepackage{pifont}  %

\colorlet{shadecolor}{yellow}
\usepackage[pdftex]{graphicx}
\graphicspath{{../pdf/}{../jpeg/}}
\DeclareGraphicsExtensions{.pdf,.jpeg,.png}

\usepackage[cmex10]{amsmath}
\usepackage{array}
\usepackage{mdwmath}
\usepackage{mdwtab}
\usepackage{eqparbox}
\usepackage{url}

\usepackage{amssymb,amsfonts}
\usepackage{algorithmic}
\usepackage{graphicx}
\usepackage{textcomp}
\usepackage{xcolor}
\usepackage{amsmath}
\usepackage{bm}
\usepackage{empheq}
\usepackage{caption}
\usepackage{subcaption}
\usepackage{xurl}
\usepackage[normalem]{ulem}
\usepackage{cancel}
\usepackage[font=small,labelfont=bf]{caption}
\usepackage{makecell}
\usepackage{amssymb}
\newcommand{\bb}[1]{{\color{black}#1}}

\def\BibTeX{{\rm B\kern-.05em{\sc i\kern-.025em b}\kern-.08em
		T\kern-.1667em\lower.7ex\hbox{E}\kern-.125emX}}
\AtBeginDocument{\definecolor{tmlcncolor}{cmyk}{0.93,0.59,0.15,0.02}\definecolor{NavyBlue}{RGB}{0,86,125}}

\def\OJlogo{\vspace{-4pt}$<$Society logo(s) and publication title will appear here.$>$}
\def\seclogo{\vspace{10pt}$<$Society logo(s) and publication title will appear here.$>$}

\begin{document}
	\receiveddate{XX Month, XXXX}
	\reviseddate{XX Month, XXXX}
	\accepteddate{XX Month, XXXX}
	\publisheddate{XX Month, XXXX}
	\currentdate{XX Month, XXXX}
	\doiinfo{XXXX.2022.1234567}
	
	\markboth{}{Author {et al.}}

	\title{ISAC: From Human to Environmental Sensing}
	\author{
		Kai Wu,
		Zhongqin Wang,
		Shu-Lin Chen,
		J. Andrew Zhang,
		and~Y. Jay Guo}

\affil{All authors are with the Global Big Data Technologies Centre, University of Technology Sydney, Australia.}
\corresp{Corresponding author: Y. Jay Guo (email: jay.guo@uts.edu.au).}

\begin{abstract}
Integrated Sensing and Communications (ISAC) is poised to become one of the defining capabilities of the sixth generation (6G) wireless communications systems, enabling the network infrastructure to jointly support high-throughput communications and situational awareness. While recent advances have explored ISAC for both human-centric applications and environmental monitoring, existing research remains fragmented across these domains. This paper provides the first unified review of ISAC-enabled sensing for both human activities and environment, focusing on signal-level mechanisms, sensing features, and real-world feasibility. We begin by characterising how diverse physical phenomena, ranging from human vital sign and motion to precipitation and flood dynamics, impact wireless signal propagation, producing measurable signatures in channel state information (CSI), Doppler profiles, and signal statistics. A comprehensive analysis is then presented across two domains: human sensing applications including localisation, activity recognition, and vital sign monitoring; and environmental sensing for rainfall, soil moisture, and water level. Experimental results from Long-Term Evolution (LTE) sensing under non-line-of-sight (NLOS) conditions are incorporated to highlight the feasibility in infrastructure-limited scenarios. 
Open challenges in signal fusion, domain adaptation, and generalisable sensing architectures are discussed to facilitate future research toward scalable and autonomous ISAC.
\end{abstract}


\maketitle

\section{Introduction}

Integrated sensing and communications (ISAC) is emerging as a hallmark feature for 6G. By combining the functionalities of wireless communications and radio sensing within a unified system, ISAC will enable 6G devices and infrastructure to not only provide connectivity but also sense the physical environment \cite{book_wu2022joint,zhang2021enabling, zhang2021overview, lu2024integrated}. At its core, ISAC aims to unify the physical-layer operations of communications and sensing by sharing spectral, temporal, and hardware resources, resulting in a system that is bandwidth- and energy-efficient, context-aware, and perceptive. This co-design reduces redundancy in RF hardware chains, enables spectrum reuse and, most importantly, makes it possible to sense the world using the ubiquitous communications network infrastructure. The applications of ISAC are only limited by our imaginations, ranging from smart cities, intelligent transport, industrial automation, digital twins and human-machine interaction to environmental sensing \cite{zhang2021enabling}.

The concept of ISAC can be rooted in dual-function radar-communication (DFRC) systems. Early research demonstrated the feasibility of embedding communication information into radar waveforms using techniques like sidelobe control, waveform diversity, and phase modulation \cite{hassanien_signaling_2016, hassanien_dual-function_2019, kai_wu2021frequency}. These pioneering DFRC methods showed that excellent communications can be achieved without degrading radar utilities, leading to architectures capable of secure \cite{kai_secure_dfrc}, embedded data transmission even under the constraints of radar operations \cite{kai_reliable_dfrc}. The concept of perceptive mobile networks (PMNs) was then introduced, extending joint communications and sensing (JSAC) capabilities to cellular networks \cite{zhang_perceptive_2021}, thereby facilitating the widespread ISAC research activities. 

ISAC has gained formal recognition in global standardisation efforts, with the ITU-R including it in the IMT-2030 vision. Kaushik et al. \cite{kaushik_toward_2024} reviewed key enabling technologies explored by 3GPP and other bodies, such as orthogonal time-frequency space (OTFS) modulation, metasurface-aided ISAC, and interference management. Their work outlined the transition of ISAC from conceptual feasibility to commercial integration, addressing physical-layer and protocol-level challenges. The expansion of ISAC across diverse domains includes significant advancements in hardware systems, such as antennas and metasurfaces.

\bb{Recent advances in antenna design have significantly contributed to the evolution of ISAC, enabling the joint fulfilment of high-capacity data transmission and fine-grained environmental perception. Massive MIMO systems, known for their spatial multiplexing and beamforming capabilities, play a foundational role in high-resolution sensing and robust communication \cite{AkroutM_2023JSAC, ZhangR_2024TWC}. Complementing this, analog multibeam antennas employing reconfigurable beamforming networks, such as Butler matrices and generalised joined coupler (GJC) matrices, allow simultaneous beam generation with low power consumption, facilitating energy-efficient and scalable ISAC deployments \cite{LiM_2025TVT, WuK_2023CM, guo2021advanced}. Simultaneous transmit and receive (STAR) antennas further advance ISAC by enabling full-duplex operation with high isolation and wide bandwidth, ideal for IoT sensing applications \cite{MaL_2023IOT}. 

Metasurface-based designs, such as space-time-coding (STC) metasurface antennas, introduce programmable control over amplitude, phase, and polarisation, enabling adaptive sensing-communication trade-offs at the physical layer \cite{wu2023universal}. In parallel, four-dimensional (4D) antenna arrays integrate spatial, polarimetric, Doppler, and temporal domains, supporting low-probability-of-intercept (LPI) radar-communication fusion for secure and covert ISAC \cite{ChenK_2022TAP}. Finally, the fluid antenna system presents a paradigm shift by enabling reconfigurable antenna positions within a liquid medium, providing dynamic diversity and spatial adaptability, particularly suited for mobility-constrained or form-factor-limited platforms \cite{NewW_2024CST}. These innovations collectively underpin the physical layer transformation required for next-generation ISAC systems, particularly in 6G and beyond.}

Alongside hardware advancements, AI is playing a central role in enabling ISAC scalability and adaptability. Wu et al. \cite{wu_ai-enhanced_2024} propose a framework for AI-enhanced ISAC, where sensing and communication components interact through shared learning modules for applications like AI-guided beam alignment, context-aware waveform selection, and neural feature extraction. Luo et al. \cite{luo_integrated_2024} furthered this by proposing a physical-layer abstraction of the world into static environments, dynamic objects, and material layers, thus supporting real-time sensing in digital twin applications. Strinati et al. \cite{strinati_distributed_2024} introduced Distributed Intelligent ISAC (DISAC), integrating semantic communications and distributed sensor coordination for low-power, high-precision situational awareness. \bb{Furthermore, ISAC-enabled 6G digital twin (DT) designs are emerging \cite{10255711, 10742564, 10944626}, which envision real-time replicas of physical environments supported by monostatic or coordinated bistatic sensing. These systems typically require synchronised transmissions, tailored waveforms, and sensing-aware protocols. These efforts represent a long-term vision of fully integrated systems with joint performance optimisation.}

\bb{It is obvious that ISAC research has spanned a wide spectrum from waveform and hardware co-design in joint radar-communication systems to deployment-ready sensing that leverages existing wireless infrastructure.  In this work, we focus on a practical ISAC scenario where bi-static sensing is performed using existing communication signals, such as 5G or WiFi transmissions, captured by uplink/downlink receivers. This setup does not assume synchronisation with the transmitter or access to transmitted waveforms, and avoids system-level modifications. It enables low-cost, scalable, and deployment-ready sensing of environmental parameters (e.g., rainfall, water level) with minimal reliance on additional infrastructure. Readers interested in broader ISAC architectures and sensing configurations are referred to recent review work \cite{zhang2021enabling, zhang2021overview, lu2024integrated}.}

Despite extensive progress in ISAC research, a critical gap remains unaddressed: the lack of a unified, comparative understanding of ISAC sensing across both human and environmental domains. Existing literature tends to fall into two categories: either centred on physical-layer technologies such as waveform design and MIMO architectures, or constrained to domain-specific applications like vehicular radar or semantic communications. In contrast, system-level sensing tasks, such as tracking human respiration using Doppler shifts or inferring rainfall intensity through signal attenuation, operate under markedly different propagation conditions, signal statistics, and performance requirements. Human ISAC sensing typically occurs in controlled indoor or near-field settings, where subtle motion and physiological signals can be extracted from phase, Doppler, and temporal patterns. Environmental ISAC sensing, on the other hand, involves outdoor, long-range scenarios characterised by material-induced attenuation, multi-hop reflections, and wide-area multipath dynamics. These fundamental differences lead to divergent feature types, hardware constraints, and deployment strategies, underscoring the need for a generalised framework that can bridge these use cases under a shared signal-processing and inference architecture.

\begin{figure*}[!t]
	\centering
	\includegraphics[width=0.9\linewidth]{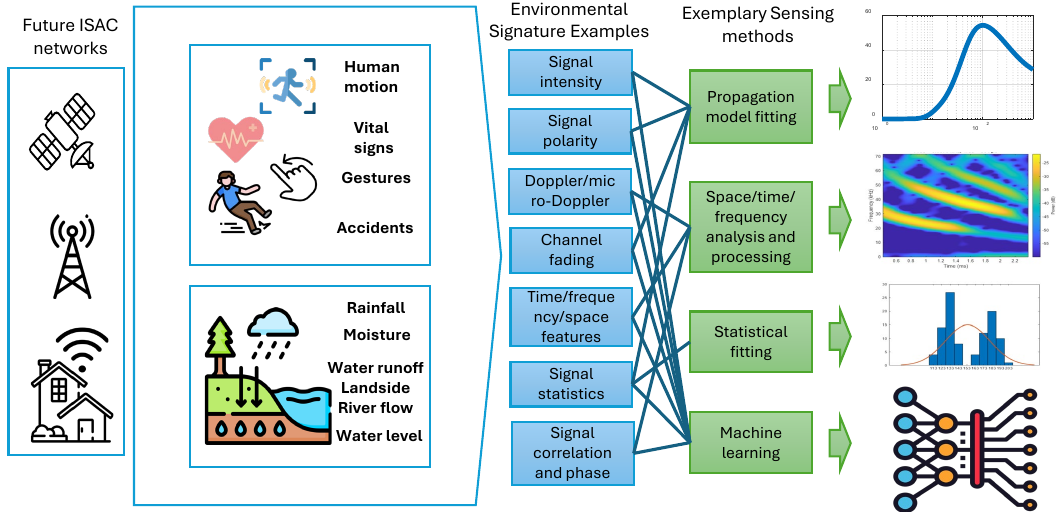}
	\caption{Conceptual framework unifying human and environmental ISAC sensing. Physical phenomena alter wireless signal propagation (e.g., through Doppler, fading, or phase shifts), producing observable signatures such as CSI fluctuations or time–frequency patterns. These are processed using propagation models, signal statistics, or learning-based inference to detect both human  and environmental activities.}
	\label{fig:isac_env_framework}
	\vspace{-1.5em}
\end{figure*}

\bb{As illustrated in Fig.~\ref{fig:isac_env_framework}, both human and environmental phenomena interact with RF signals via common mechanisms like scattering, absorption, and Doppler shifts. However, their underlying characteristics differ. Human sensing often targets fine-grained, motion-induced variations (e.g., micro-Doppler), while environmental sensing, particularly for weather and water-level monitoring, involves large-scale, distributed changes in signal propagation, such as attenuation by rainfall or reflections from water surfaces. Interference sources also vary. Body shadowing and orientation dominate in human sensing, whereas terrain, weather, and multipath effects are key in environmental contexts. Despite these differences, both domains yield identifiable signal signatures in channel state information (CSI), received signal strength indicator (RSSI), or similar metrics. ISAC provides a unified physical framework for decoding such signatures to support diverse sensing objectives.}

This paper addresses a critical gap in ISAC research by providing the first cross-domain, application-layer review of ISAC-enabled sensing, focusing on real-world feasibility and experimental validation. It consolidates fragmented research efforts by comparing signal features, processing methodologies, and inference strategies across human and environmental sensing domains. This unified framework accommodates diverse channel conditions, infrastructure constraints, and use cases, laying the foundation for developing generalizable ISAC sensing systems. 

The paper consists of two main parts. The first part reviews ISAC for human applications, including tracking, vital sign monitoring, and activity recognition. The second part delves into environmental sensing using ambient RF signals, including rainfall, soil moisture, and water level sensing, validated through experiments with LTE and WiFi signals under non-line-of-sight (NLOS) conditions. In each aspect, representative works are critically reviewed, feature extraction and modelling approaches are summarised, and shared challenges are identified with future visions provided.

The remainder of the paper is structured as follows. Section~II reviews the physical propagation model for ISAC-enabled sensing. Additionally, it highlights key communication metrics for sensing applications. Section~III presents recent advances in ISAC-based human sensing, covering typical applications such as localisation and tracking, vital sign estimation, and activity recognition. Section~IV provides a comprehensive analysis of environmental sensing use cases. It synthesises literature and field results in rainfall sensing, soil moisture estimation, and water level monitoring related to ISAC. 
It also provides a water sensing case study based on our preliminary experimental results. 
Section V concludes the paper by summarising open challenges and highlighting future research directions.

\section{ISAC Fundamentals}
\label{sec:isac_principles}
To begin with, we outline the fundamentals and core principles of ISAC, focusing on signal modelling, key signal metrics, and their application to both human and environmental sensing.

\subsection{Wireless Channel Propagation Model}
Most ISAC studies so far are based on orthogonal frequency-division multiplexing (OFDM) or alike multi-carrier waveforms, such as OTFS. OFDM has wide applications in modern wireless communications systems including WiFi, cellular and satellite. Hence, our signal model descriptions
here are based on OFDM. After channel estimation in OFDM systems, a so-called channel frequency response (CFR) can be obtained, which depicts the interactions of wireless signals with propagation environments. 
In particular, CFR describes how the channel varies with frequency and is influenced by different factors such as delay, Doppler shift, and the Angle of Arrival (AoA) and Angle of Departure (AoD) \cite{zhang2022integration, kai_wu2024sensing, wang2024passive}. In a typical multi-antenna system, the CFR from the $i_t$-th transmitting antenna and the $i_r$-th receiving antenna over the $j$-th subcarrier and $k$-th OFDM symbol can be expressed as:
\begin{equation}
\resizebox{\linewidth}{!}{$
\begin{aligned}
    H_{i_t,i_r,j,k} =& \beta_{k} \underbrace{e^{-J [2\pi \left(\Delta f_j \Delta \tau_k^{\text{TO}} + \Delta f_k^{\text{CFO}} k T_s \right)+ \phi_{i_t,i_r}]}}_{\text{Phase Offsets}} \times \\
    &\sum_{l=1}^{N_p} \Big[ \rho_{i_t,i_r,j,k}[l] 
    \underbrace{e^{-J 2\pi \Delta f_j \tau[l]}}_{\text{Delay}} 
    \underbrace{e^{-J 2\pi f^D[l] k T_s}}_{\text{Doppler}} \times \\
    & \underbrace{e^{-J 2\pi \frac{f_c}{c} (i_r-1) \Delta d_{r} \sin\theta^{r}[l]}}_{\text{AoA}}
    \underbrace{e^{-J 2\pi \frac{f_c}{c} (i_t-1) \Delta d_{t} \sin\theta^{t}[l]}}_{\text{AoD}} \Big],
\end{aligned}
$}
\label{equation1}
\end{equation}
where $\beta_k$ is the power variation factor, caused by effects such as automatic gain control (AGC) or variations in transmission power; $\Delta \tau^{\text{TO}}$ and $\Delta f^{\text{CFO}}$ denote the timing offset (TO) and carrier frequency offset (CFO) caused by clock asynchrony; $\phi_{i_t,i_r}$ is the hardware-induced offset, such as differences between antennas, cables, and RF chains; $\rho_{i_t,i_r,j,k}$ is the path loss; $\Delta f_j$ is the frequency of the $j$-th subcarrier; $\tau$ is the propagation delay; $f^D$ is the Doppler shift; $T_s$ is the OFDM symbol interval; $\theta^r$ and $\theta^t$ are the AoA and AoD, respectively; $\Delta d_r$ and $\Delta d_t$ represent the spacing between antennas at the receiver and transmitter; $f_c$ is the centre carrier frequency; $c$ is the speed of light.

\bb{The CFR model in \eqref{equation1} supports both human and environmental sensing, but key components differ. Human sensing involves localised, periodic changes in, e.g., $f^D[l]$, $\tau[l]$, and the set of $l$ over a small set of $k$, often requiring fine resolution in $\theta^r[l]$ and $\theta^t[l]$. Environmental sensing, in contrast, exhibits broader and slower changes. Rainfall or water level shifts affect $\rho_{i_t,i_r,j,k}[l]$, $\tau[l]$, and $\beta_k$ across $j$ and $k$. NLOS reflections from terrain or water introduce wide angular spread in $\theta^r[l]$ and $\theta^t[l]$, favouring statistical features over instantaneous inference.}

\subsection{Communication Metrics as Sensing Proxies}
In commercial wireless communication systems, various physical-layer measurements are available, thus enabling accurate sensing without hardware modification. In the following, we focus on commonly used metrics, including signal strength indicators and CSI.

\subsubsection{Signal Strength and Quality Indicators}
Signal strength features, including RSSI, Reference Signal Received Power (RSRP), Reference Signal Received Quality (RSRQ), Signal-to-Interference-plus-Noise Ratio (SINR), are among the most accessible and widely used metrics in wireless sensing \cite{shin2023lte, 10556755, miao2025wi}. These features are natively supported by most commercial devices and are easy to extract without specialised hardware or drivers. For instance, RSSI can be directly accessed from smartphones, WiFi routers, and IoT devices via standard application programming interfaces (APIs). Likewise, these indicators are routinely reported by cellular modems and baseband processors in 4G LTE and 5G NR terminals as part of network measurement reports. 

While these indicators provide a low-cost and readily accessible entry point for basic presence detection and coarse environmental dynamics estimation, their inherent limitations significantly restrict their effectiveness in fine-grained sensing tasks. First, their values are typically coarse, quantised, and often averaged over multiple time and frequency resource blocks, resulting in low sensing resolution. Furthermore, these indicators lack the ability to resolve individual multipath components, making it difficult to distinguish target-induced signal changes from background clutter and static reflections. As a result, sensing models trained solely on such features are often environment-dependent, exhibiting poor generalisation performance across different deployment scenarios. Despite their accessibility, these signal strength and quality indicators fall short of meeting the resolution and robustness requirements of fine-grained sensing systems.

\subsubsection{Channel State Information}
To address the limitations of coarse signal strength indicators, recent research has shifted toward using CSI, which captures fine-grained channel characteristics by providing amplitude and phase measurements for each subcarrier. CSI can be extracted from 4G/5G signals. In LTE systems, CSI is typically estimated from cell-specific reference signals (CRS) through frame synchronisation and channel estimation, and supported by open-source tools such as srsLTE \cite{gomez2016srslte} and OpenLTE \cite{nikaein2014openairinterface}. In WiFi-based sensing, CSI has been extensively studied, with several open-source tools, such as the Intel 5300 CSI Tool \cite{halperin2011tool}, Atheros CSI Tool \cite{xie2015precise}, and Nexmon CSI \cite{nexmon:project, 10.1145/3349623.3355477}, enabling CSI extraction from commodity hardware. Access to raw CSI data remains restricted in many commercial WiFi chipsets, posing challenges for widespread ISAC deployment. Compared to signal strength indicators, CSI offers higher spectral resolution and better multipath separability, making it ideal for developing fine-grained and robust ISAC applications.

\section{ISAC for Human Sensing}
The ability to sense human presence, motion, and physiological signals in a contactless, device-free manner is highly desirable for healthcare, smart environments, and public safety. ISAC-based sensing offers a passive, privacy-preserving alternative to wearable and cameras by exploiting the variation of signals from existing wireless infrastructure. In the following, we will focus on key human sensing technologies in ISAC systems, including CSI preprocessing, feature extraction, location-refined sensing, and deep learning. We then present case studies to illustrate their practical effectiveness.

\subsection{Clock Asynchronism Removal Technologies}
Despite the high resolution and rich channel characteristics provided by CSI, its application in ISAC systems---particularly bi-static architectures where the transmitter and receiver are separately deployed without a unified clock source---remains technically challenging. In such transceiver setups, CSI is affected by the TO and CFO, resulting in time-varying random phase shifts across CSI measurements. These distortions severely hinder the extraction of meaningful motion-induced features. To address this challenge, a variety of techniques \cite{zhang2022integration, 10623407, wu2023low, zhao2023multiple} have been proposed from various domains. 

\subsubsection{CACC} 
Cross-Antenna Cross-Correlation (CACC) \cite{li2017indotrack, qian2018widar2} computes conjugate multiplication between Rx antenna pairs to remove random phase offsets while preserving linear signal relationships. However, due to similar antenna power levels, Doppler mirroring may occur, making it difficult to determine motion direction. Recent works have proposed enhanced CACC methods for random phase offset removal. WiDFS \cite{wang2023single} and WiDFS 2.0 \cite{wang2024passive} introduce a variant of CACC for multi-antenna setups, which constructs differential terms using static signal components across antenna pairs rather than relying solely on cross-correlation. A linear transformation is then applied to suppress Doppler mirroring while preserving the linearity of signal components. This approach offers low computational complexity and is applicable to both single-target and multi-target scenarios.

\subsubsection{CASR}
Cross-Antenna Signal Ratio (CASR) \cite{feng2021lte, li2022csi, 10678871, ni2023uplink} uses CSI ratio between antennas to remove automatic gain control and Doppler ambiguity. Yet, its performance degrades when multiple Doppler components occur, and the resulting non-linear form complicates delay and AoA estimation.

\subsubsection{Reference Signal} 
Some methods \cite{meneghello2022sharp, pegoraro2024jump} exploit subcarrier frequencies to construct single-antenna phase references. While requiring sufficient bandwidth, they offer good hardware compatibility but also suffer from Doppler mirroring under narrowband conditions. Some approaches \cite{ dong2024signal} construct reference signals from the AoA domain using multi-antenna arrays. However, antenna hardware diversity can compromise the accuracy of the reference signal construction.

A comprehensive comparison of CSI random phase removal methods can be found in \cite{wu2024sensing}, which systematically evaluates their effectiveness, limitations, and applicability across different ISAC scenarios.

\subsection{Multi-Domain Feature Extraction}
With the intended signal captured, the next step is to extract multi-domain features, including Doppler, delay, and AoA across the temporal, spectral, and spatial dimensions, respectively. Due to the phase correction, these features are computed relative to a reference signal path. In most existing works, the Line-of-Sight (LOS) path between the fixed transmitter and receiver with strong and stable power is commonly used as the reference. Within the strongest signal component, the Doppler shift reflects motion induced purely by human motion. The delay represents the difference in path length between the NLOS reflection caused by the human body and the LOS path. And the AoA represents the angular difference between the direct Tx-Rx path and the human reflection path. However, some existing approaches \cite{li2017indotrack, niu2022rethinking} extract only Doppler features to achieve human tracking. Such Doppler-based methods often suffer from accumulated trajectory errors over time and typically require multiple Tx-Rx pairs for real velocity estimation. This increases deployment cost and limits scalability in practical systems. In contrast, the multi-dimensional parameter estimation aims to jointly extract these features and can operate with a single-input multiple-output (SIMO) configuration (a 1Tx-3Rx setup). The following mainly focuses on the multi-dimensional approaches for fine-grained human perception across different signal domains.

\subsubsection{Joint Optimisation-Based Feature Extraction}
Widar2.0 \cite{qian2018widar2} and mD-Track \cite{xie2019md} perform joint parameter estimation using maximum likelihood estimation (MLE). These methods typically rely on CACC-based pre-processing and construct a likelihood function over the three-dimensional parameter space. While they achieve effective localisation through iterative refinement, they often suffer from high computational complexity and require precise initial guesses, which may limit their real-time applicability in dynamic environments.

\subsubsection{Sequential Processing-Based Feature Extraction}
To ensure real-time performance in human sensing, WiDFS2.0 \cite{wang2024passive} proposes a lightweight multi-dimensional feature extraction method. It first applies the CACC-variant technique to eliminate random phase and Doppler mirroring. This yields a 3D CSI cube indexed by time, subcarriers, and antenna pairs. A beamforming-based feature extraction process is presented as follows:

\begin{itemize}
\item \textit{Doppler FFT.} WiDFS2.0 applies a Doppler fast Fourier transform (FFT), exploiting its high resolution to isolate moving targets in the Doppler domain. Doppler typically exhibits higher resolution than delay and AoA under communication system constraints, allowing for better separation of moving targets.

\item \textit{Delay MVDR.} For each Doppler bin, it performs delay-domain Minimum Variance Distortionless Response (MVDR) beamforming using antenna-pair snapshots.

\item \textit{AoA FFT.} For each Doppler-Delay bin, it uses MVDR weights to each antenna pair, followed by an AoA FFT to estimate the spatial direction. This process yields a complete Doppler-Delay-AoA 3D feature map. The beamforming operation in the delay and spatial domains  enhances resolution and effectively suppresses sidelobes.
\end{itemize}

\subsection{Location-Based ISAC Human Sensing}
After the multi-domain feature extraction process, a 3D feature tensor across the Doppler-Delay-AoA dimensions, along with residual signal components within each bin, can be obtained. On this basis, ISAC-based human sensing systems typically perform three core sensing tasks: human localisation and tracking, vital sign monitoring, and activity recognition. In the following, we first introduce target localisation and tracking, and then describe how the estimated position can serve as a spatial filter to focus on target regions in vital sign and activity sensing tasks, thereby improving sensing accuracy and robustness. To the best of our knowledge, most existing ISAC methods make limited use of position information as an effective filter in sensing tasks.

\subsubsection{Localisation and Tracking via Feature Point Clouds}
The work \cite{wang2024passive} designs a real-time passive human tracking framework based on Doppler-Delay-AoA features. The system integrates 2D Constant False Alarm Rate (CFAR)-based target detection and an Extended Kalman Filter (EKF)-based tracking to achieve accurate and low-latency tracking. \textit{(1) Object Detection via 2D CFAR.} The 3D Doppler-Delay-AoA feature tensor can be compressed into a 2D Doppler-AoA heatmap, and apply a 2D CFAR detection algorithm to identify potential targets. Since the Doppler and AoA domains offer relatively high resolution compared to the delay domain, this 2D projection enables accurate target detection. Each detected bin yields a point cloud, consisting of delay, AoA, Doppler, and SNR, where the SNR can be used to indicate the likelihood of a true target. \textit{(2) Real-time Tracking via EKF.} The EKF can be used to achieve tracking of human targets with low computational overhead. For more complex multi-target scenarios, Joint Integrated Probabilistic Data Association (JIPDA) can be employed to improve robustness against clutter and closely spaced targets. Due to the inherently low resolution in the delay domain, the weighted joint optimisation is performed across point clouds collected over multiple short-time windows, where SNR is used to weight the contribution of each detection. This allows more reliable trajectory initialisation and update. The complete EKF tracking process includes the following steps:

\begin{itemize}
\item \textit{Track Initialisation and Coordinate Transformation.} To initiate a new track, the delay and AoA values corresponding to each detected target point are extracted. As these values are computed relative to a known reference path (typically the LOS path), they are transformed into absolute coordinates using the known positions of the transmitter and receiver. This yields the initial position of the target in Cartesian space. The initial state of each track includes the target's position, velocity, and acceleration, based on a constant-acceleration motion model, which assumes the target moves with approximately uniform acceleration over short time intervals.

\item \textit{Track Prediction.} At each time step, the EKF predicts the future state of each active track. This includes forecasting the target's position, velocity, and acceleration based on the previous state. The prediction step allows the tracker to estimate the most probable location of each target in the next frame, providing a basis for subsequent data association. This is particularly important when dealing with temporary occlusion or cluttered environments, where direct observations may be intermittent or noisy.

\item \textit{Point-to-Track Association.} Once new detections are available, the tracker attempts to associate them with existing tracks. In simple scenarios, this can be done using Euclidean distance. In complex environments, a more robust Mahalanobis distance can be used, which considers multiple features including position, Doppler, and SNR. Those detections that fall within a defined gating region are assigned to tracks.

\item \textit{Assigned Track Update.} If a valid detection is assigned to a track, the EKF updates its state estimate. The new measurement is combined with the predicted state to refine the track's estimate of the target's position, velocity, and acceleration. This step improves tracking accuracy and corrects accumulated prediction errors.

\item \textit{Unassigned Track Update.} Tracks that do not receive a matching detection in a given frame are not immediately discarded. Instead, their states can be updated based on motion prediction alone and marked as ``tentative". If a track continues to be unassigned over several frames, it is eventually removed.

\item \textit{Creating New Tracks.} Any new detections that are not associated with existing tracks are treated as potential new targets. These are used to initialise new tracks, which enter the tracker as tentative candidates. Tracks are promoted to confirmed status once they are consistently detected over multiple frames.

\item \textit{Maintaining Active Tracks.} The tracker checks the status of all tracks. Confirmed tracks are maintained as long as they continue to receive updates or valid predictions. Tentative or low-confidence tracks are removed if they fail to meet confirmation or persistence criteria. Additionally, in static environments where no human motion is present, the system should theoretically observe no active tracks. Therefore, the absence of any initialised or sustained tracks can be used as an indicator that no moving targets are currently present, providing a lightweight mechanism for motion presence detection.
\end{itemize}

\subsubsection{Location-based Vital Sign Estimation}
Most existing approaches \cite{soto2022survey} directly utilise CSI phase variations to detect a target's vital signs. However, the Doppler-Delay-AoA features can be employed as a spatial-temporal filter to effectively isolate signal components associated with the estimated position of the target. The phase information of the residual CSI signals in the selected bin is then used, as it is highly sensitive to subtle micro-motions of the human body. This enables non-contact estimation of vital signs such as respiration and heartbeat. The underlying principle is based on the periodic movement of the chest and heart, which induces small changes in the propagation path length and, consequently, the phase of the received signal. However, for communication signals with lower operating frequencies (e.g., 2.6 GHz LTE or 5 GHz WiFi), the longer wavelengths (approximately 11.5 cm and 6 cm, respectively) enable the detection of larger periodic motions such as breathing, but are somewhat less effective at capturing subtle movements like heartbeats. The process consists of the following key steps:

\begin{itemize}
\item \textit{Static Target Localisation and Phase Extraction.} For stationary human targets, the dominant motion typically arises from the rhythmic expansion and contraction of the chest. To isolate such subtle movements from environmental clutter, we leverage the Doppler dimension to identify low-velocity components corresponding to static human reflections. A longer time window is required to improve Doppler resolution and suppress noise. Based on this, we identify the target's location bin from the estimated Doppler-Delay-AoA feature cube, selecting the strongest static reflection associated with the human body. Within the selected bin, we extract the phase of the residual CSI signals over time.

\item \textit{Phase Differencing.} As the human body cannot remain perfectly stationary, slow drifts and body sway introduce phase fluctuations unrelated to respiration or heartbeat. To suppress these motion impacts, we compute the differential phase, which highlights  chest and heart micro-motions while suppressing low-frequency trends.

\item \textit{Filtering and Frequency Estimation.} The differential phase signal is then passed through two bandpass filters to isolate the typical frequency ranges of respiration (e.g., 0.1-0.5 Hz) and heartbeat (e.g., 0.8-2 Hz). We subsequently apply frequency estimation methods such as the FFT or high-resolution spectral estimators (e.g., MUSIC, ESPRIT) to extract the dominant frequency components, enabling accurate vital sign estimation.
\end{itemize}

\begin{figure*}[t]
    \centering
\begin{subfigure}{0.245\linewidth}
        \centering
        \includegraphics[width=\textwidth]{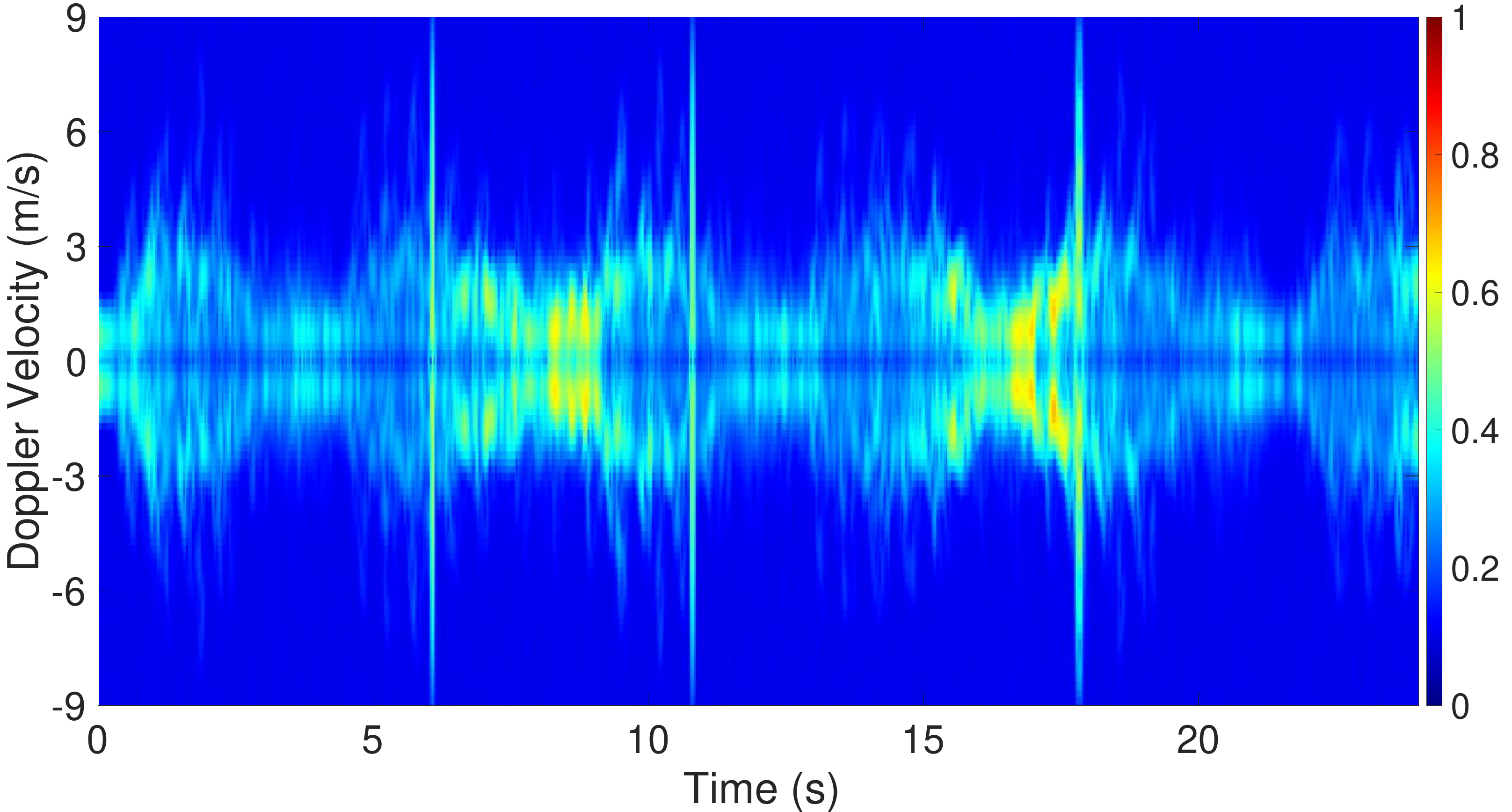}
        \subcaption{Raw CACC}
        \label{Fig_raw_cacc}
    \end{subfigure}
    \begin{subfigure}{0.245\linewidth}
        \centering
        \includegraphics[width=\textwidth]{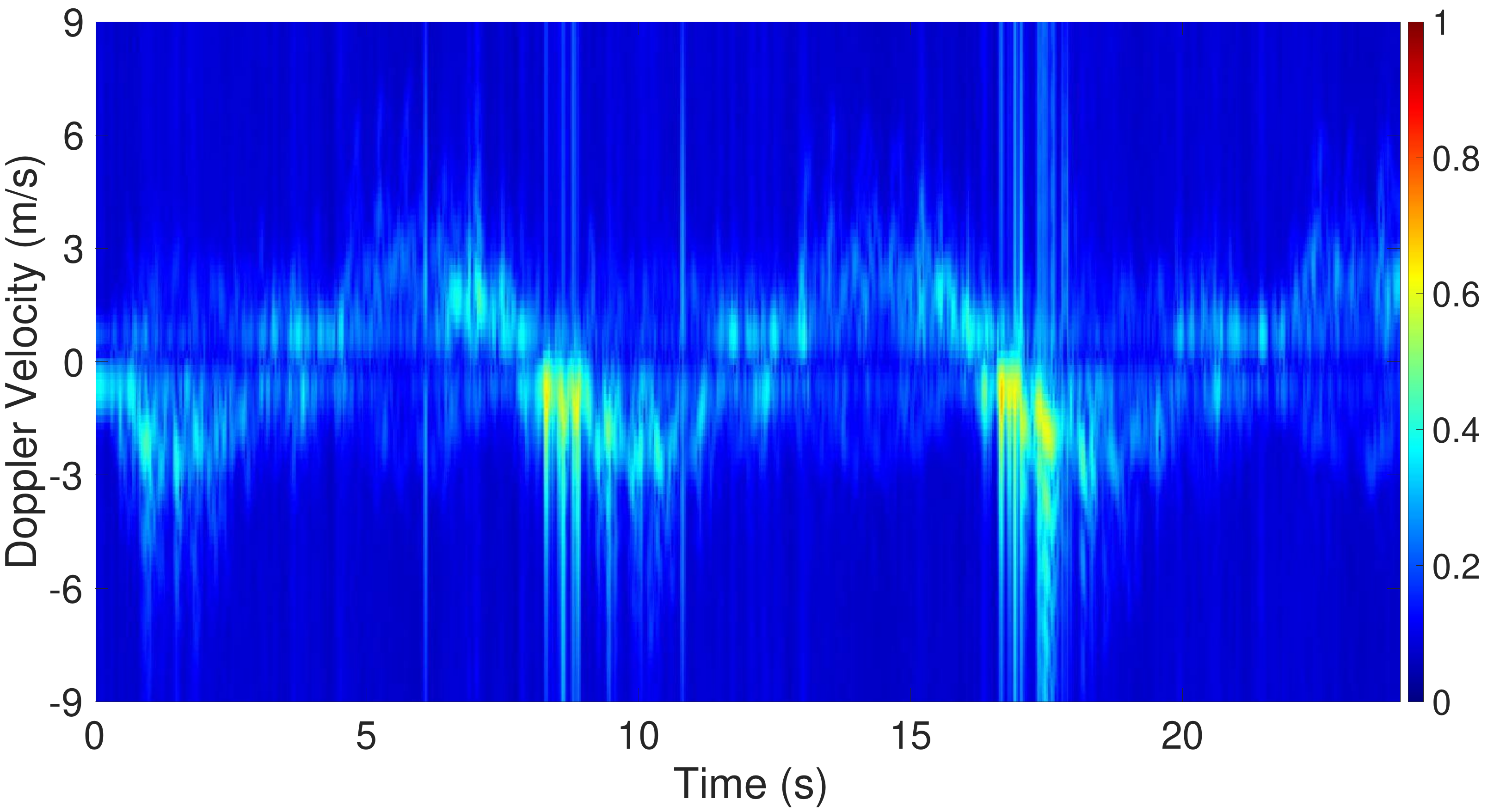}
        \subcaption{CASR}
        \label{Fig_csiratio}
    \end{subfigure}
    \begin{subfigure}{0.245\linewidth}
        \centering
        \includegraphics[width=\textwidth]{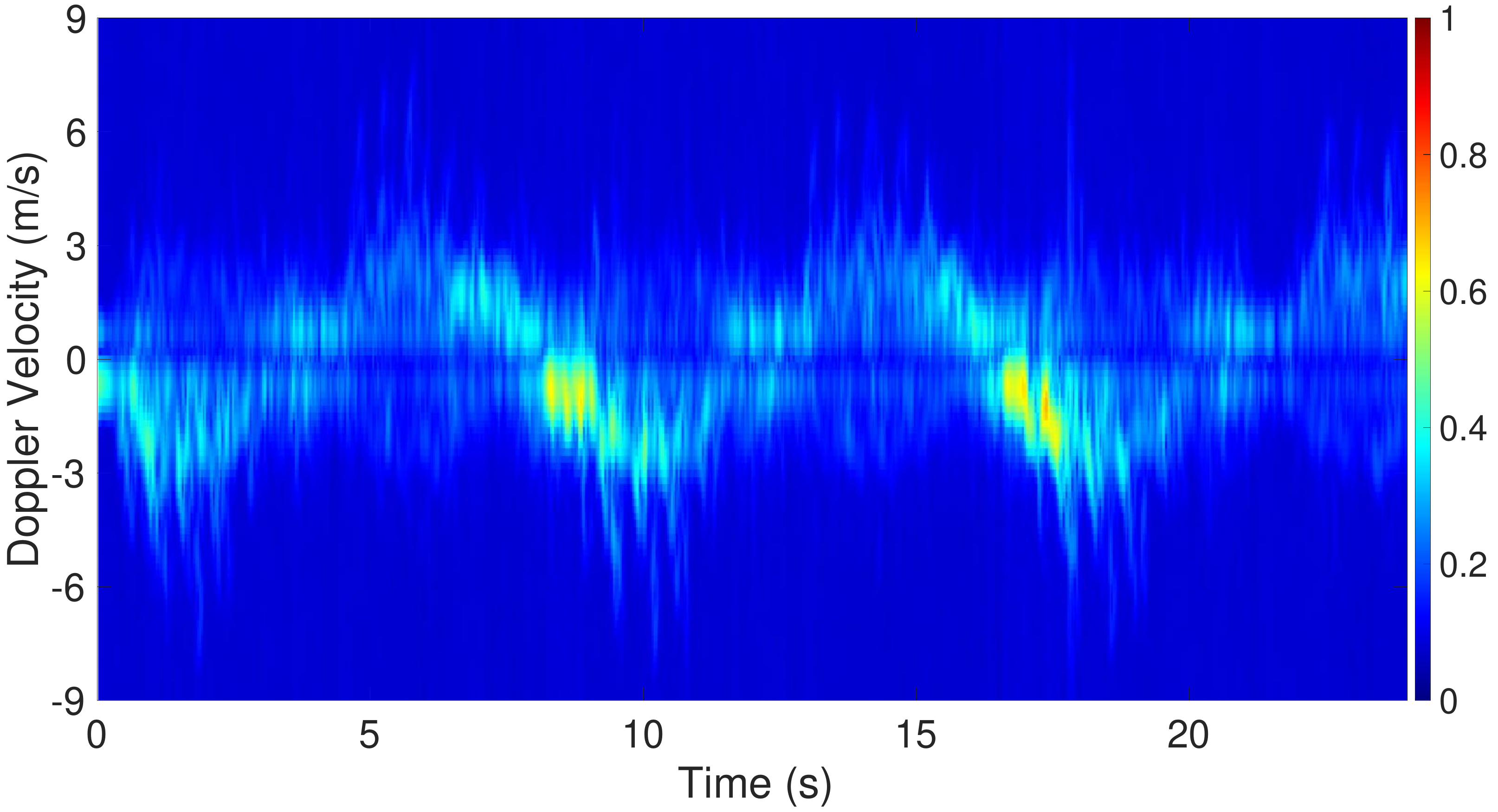}
        \subcaption{CACC-variant}
        \label{Fig_cacc_variant}
    \end{subfigure}
    \begin{subfigure}{0.245\linewidth}
        \centering
        \includegraphics[width=\textwidth]{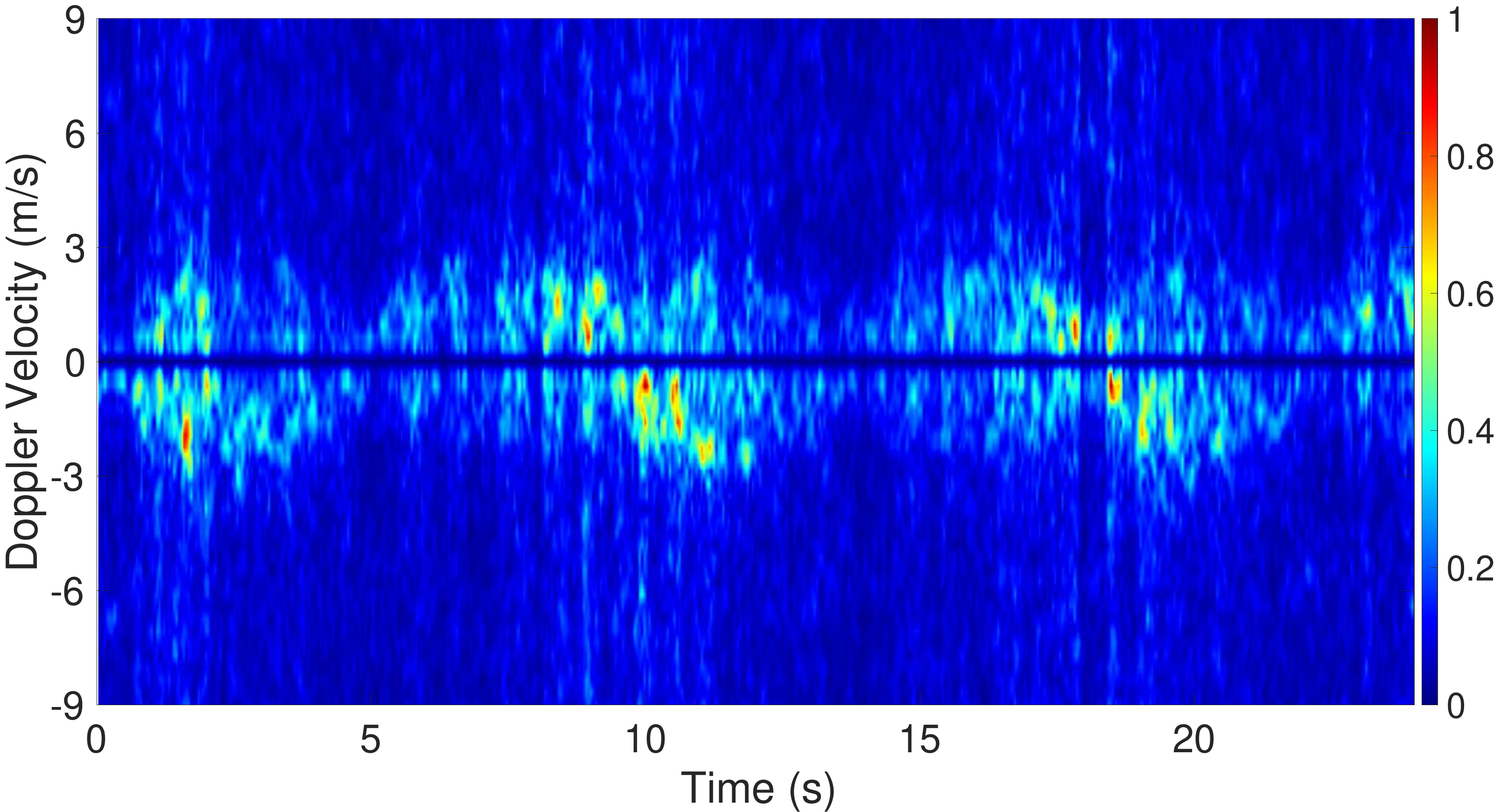}
        \subcaption{Single-antenna}
        \label{Fig_single_antenna}
    \end{subfigure}
    \caption{\bb{Doppler-time heatmaps demonstrating the effectiveness of random phase removal methods based on 3.1 GHz LTE signals.}}
    \label{Fig_phase_removal}
\end{figure*}

\subsubsection{Location-based Activity Recognition} 
For activity recognition, we can aggregate high-resolution Doppler signatures over multiple time windows at the target's estimated location to construct Doppler features, enabling fine-grained characterisation of human movements. Building upon this, additional motion cues such as gait patterns \cite{yan2025pushing} can also be extracted, which further support user identification tasks \cite{yang2025environment}. The human body is a non-rigid structure, and different activities involve different body parts moving at various speeds and directions, which induce distinct patterns in the Doppler domain-referred to as the micro-Doppler effect \cite{heo2025millimeter}. While body movement also affects the delay and AoA dimensions, the limited bandwidth and antenna amount in practical ISAC systems result in low resolution, making it difficult to resolve fine-grained limb movements in delay and spatial domains. In contrast, the Doppler domain offers higher resolution, making it a primary feature for activity sensing. Importantly, the Doppler patterns generated by the same activity (e.g., walking, waving, falling) tend to be similar and repeatable. This characteristic enables the application of machine learning and deep learning models to classify activities or gestures. The process consists of the following three key stages:

\begin{itemize} 
\item \textit{Micro-Doppler Feature Extraction.} For each time window, we extract the 1D Doppler signature from the Doppler-Delay-AoA feature cube by selecting the Doppler profile at the target's identified delay and AoA bin. Since human activities are continuous in time, we aggregate Doppler profiles across consecutive time windows to form a micro-Doppler representation, which captures dynamic patterns of body movement. Depending on the applications, the window duration and overlap size can be adjusted to obtain finer temporal resolution or better frequency stability.

\item \textit{Temporal Segmentation.} A key challenge in activity classification is how to segment continuous behaviours into meaningful temporal units. We propose treating this as an automatic segmentation task, analogous to voice activity detection in speech processing, and are currently exploring the use of pre-trained speech segmentation networks to perform unsupervised activity boundary detection.

\item \textit{Activity Classification.} Identical activities typically produce similar Doppler patterns, which can be classified using template matching, classical machine learning, or deep learning approaches. However, several real-world factors introduce significant variability into the micro-Doppler signatures, including differences in user position, body morphology, motion style, and transceiver placement. These variations can impact the Doppler magnitude, frequency dispersion, and the temporal evolution of features, thereby affecting classification accuracy. Therefore, effective activity recognition requires not only the extraction of high-resolution and robust micro-Doppler features, but also the design of models with strong generalisation capability.
\end{itemize}

\subsection{Data-Driven Learning Human Sensing}
In recent years, data-driven deep learning techniques have played an increasingly prominent role in ISAC-enabled human sensing. Advances in convolutional neural networks, recurrent architectures, and transformer-based models have enabled the extraction of high-level semantic representations from raw signal inputs. These models have demonstrated promising performance in various sensing tasks. More recently, the integration of large language models (LLMs) and foundation models into sensing applications has opened new opportunities. These foundation model-driven sensing offers enhanced reasoning, better generalisation, and cross-domain adaptability, marking a new frontier in ISAC research \cite{LLM_ISAC}.

\subsubsection{Unrefined Signal Representation} 
Despite growing interest in deep learning for ISAC, most existing studies \cite{yang2023sensefi, Yan_2024_CVPR, luo2024vision} rely on raw CSI data and thus suffer from random phase offsets, or apply relatively coarse-grained feature extraction prior to learning. For example, many works \cite{tang2023mdpose} remove random phase and directly perform Doppler FFT to obtain spectrogram-like features, which are then processed by well-designed neural networks. While these approaches have yielded encouraging results, they tend to primarily focus on network architecture design, with less emphasis on the physical characteristics and interpretability of the input features. As a result, the generalisation of these models across different environments and deployment conditions remains limited.

\subsubsection{Refined Signal Representation} 
The learning process should begin with the extraction of meaningful and interpretable features guided by signal models, followed by deep learning-based classification or regression. Unlike image or text inputs, wireless sensing signals are governed by well-defined physical models. This structure enables the extraction of high-level features that carry physical meaning prior to learning. These features not only enhance model interpretability but also facilitate better generalisation. For example:

\begin{itemize}
\item \textit{Localisation and Tracking.} One can extract candidate targets from the Doppler-Delay-AoA feature cube using interpretable methods such as 2D CFAR, which is grounded in radar signal processing theory. These extracted point clouds can be then processed by learning-based models \cite{wang2023human, fan2024enhancing} for trajectory association and  prediction.

\item \textit{Vital Sign Estimation} One can analyse the phase variations corresponding to chest and heart movement. These features are directly tied to physiological activity and can be further enhanced using ECG-inspired temporal deep learning models to improve respiration and heart rate detection accuracy.

\item \textit{Activity Recognition.} One can construct micro-Doppler features from the time evolution of Doppler signatures. These patterns are closely aligned with body kinematics and allow us to incorporate prior knowledge from radar-based sensing into the ISAC framework.
\end{itemize}

\begin{figure*}[t]
	\centering
	\begin{subfigure}{0.325\linewidth}
		\centering
		\includegraphics[width=\textwidth]{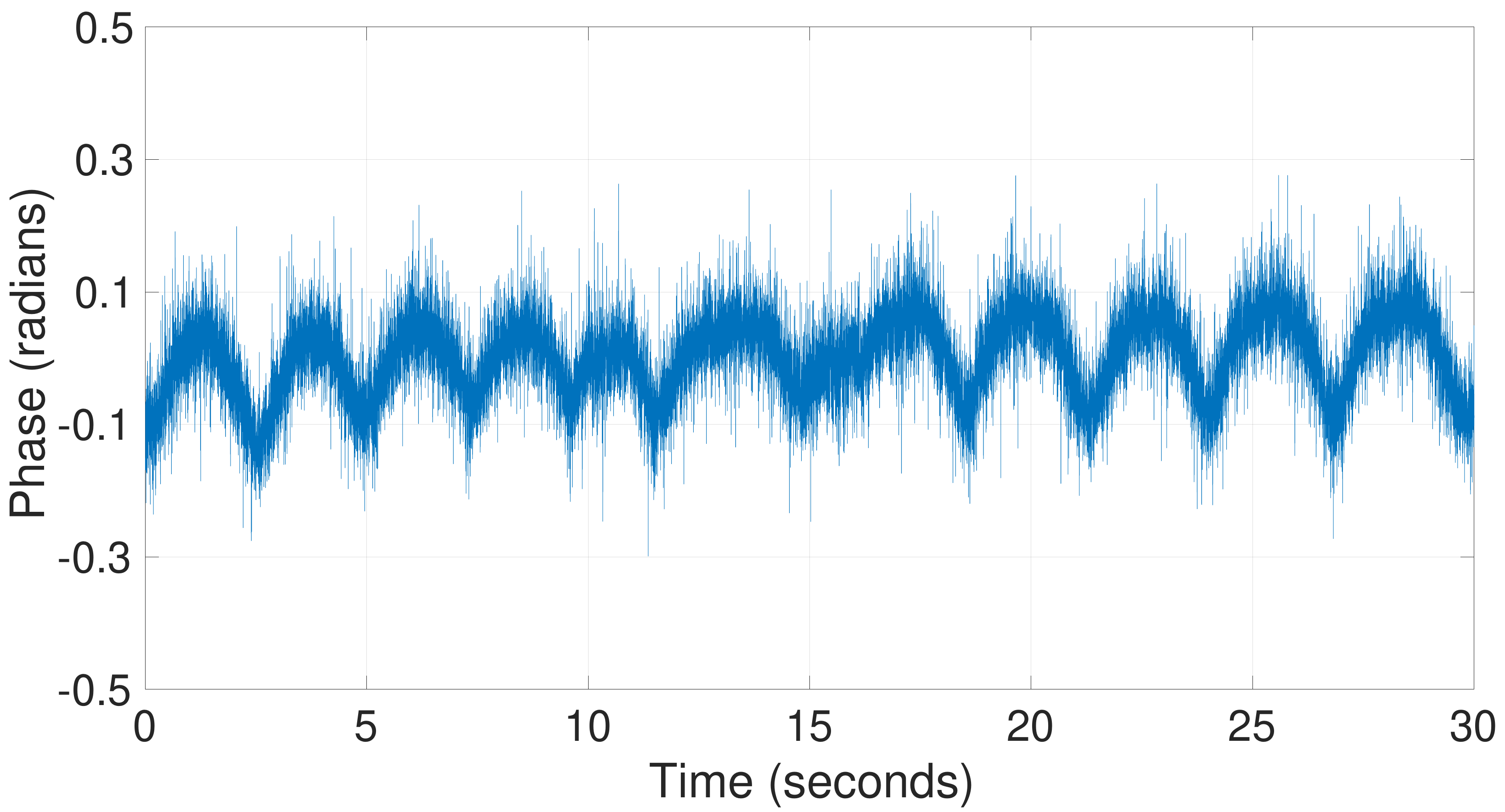}
		\subcaption{Raw phase feature}
	\end{subfigure}
	\begin{subfigure}{0.325\linewidth}
		\centering
		\includegraphics[width=\textwidth]{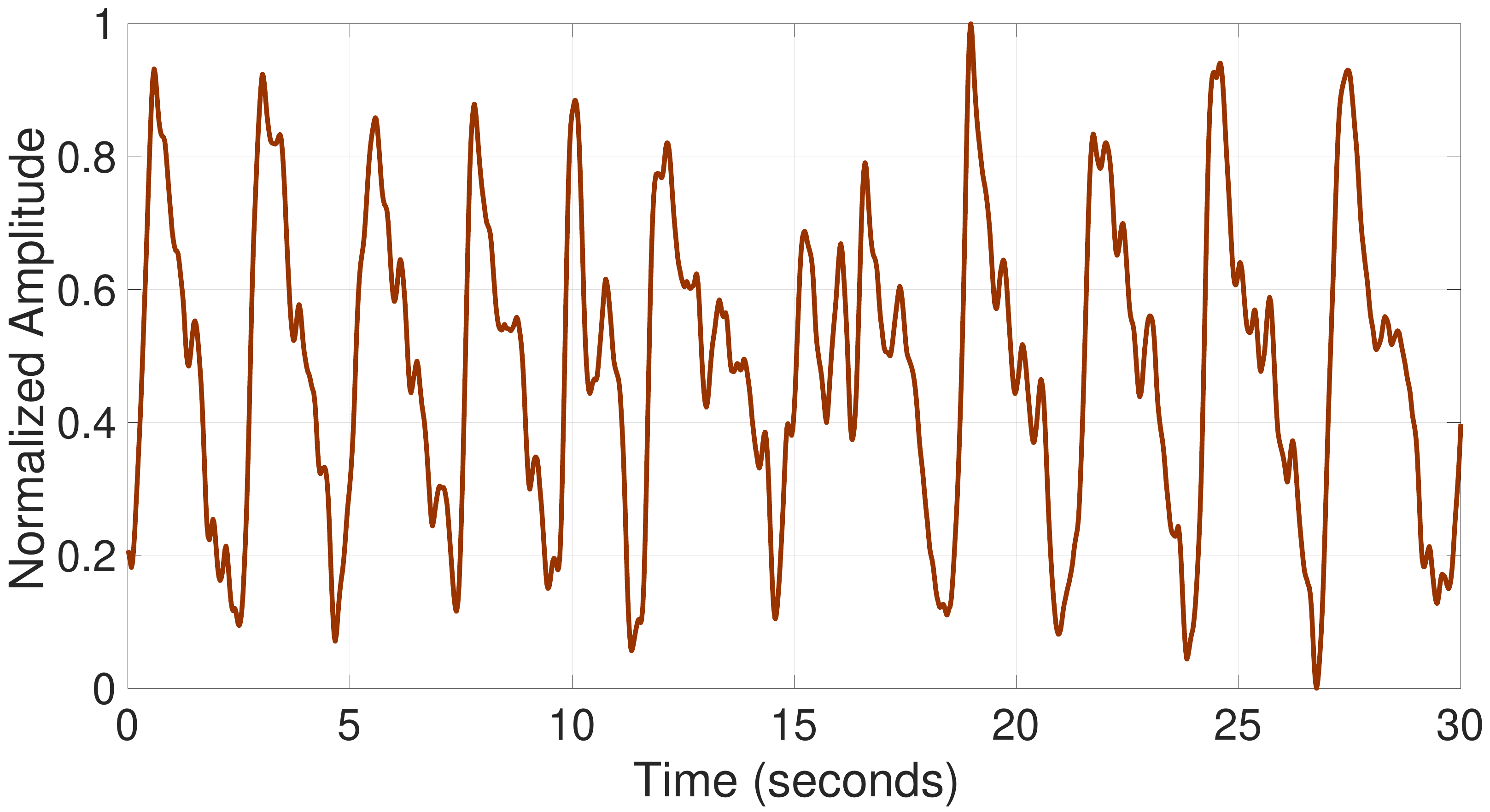}
		\subcaption{Extracted respiration waveform}
	\end{subfigure}
	\begin{subfigure}{0.325\linewidth}
		\centering
		\includegraphics[width=\textwidth]{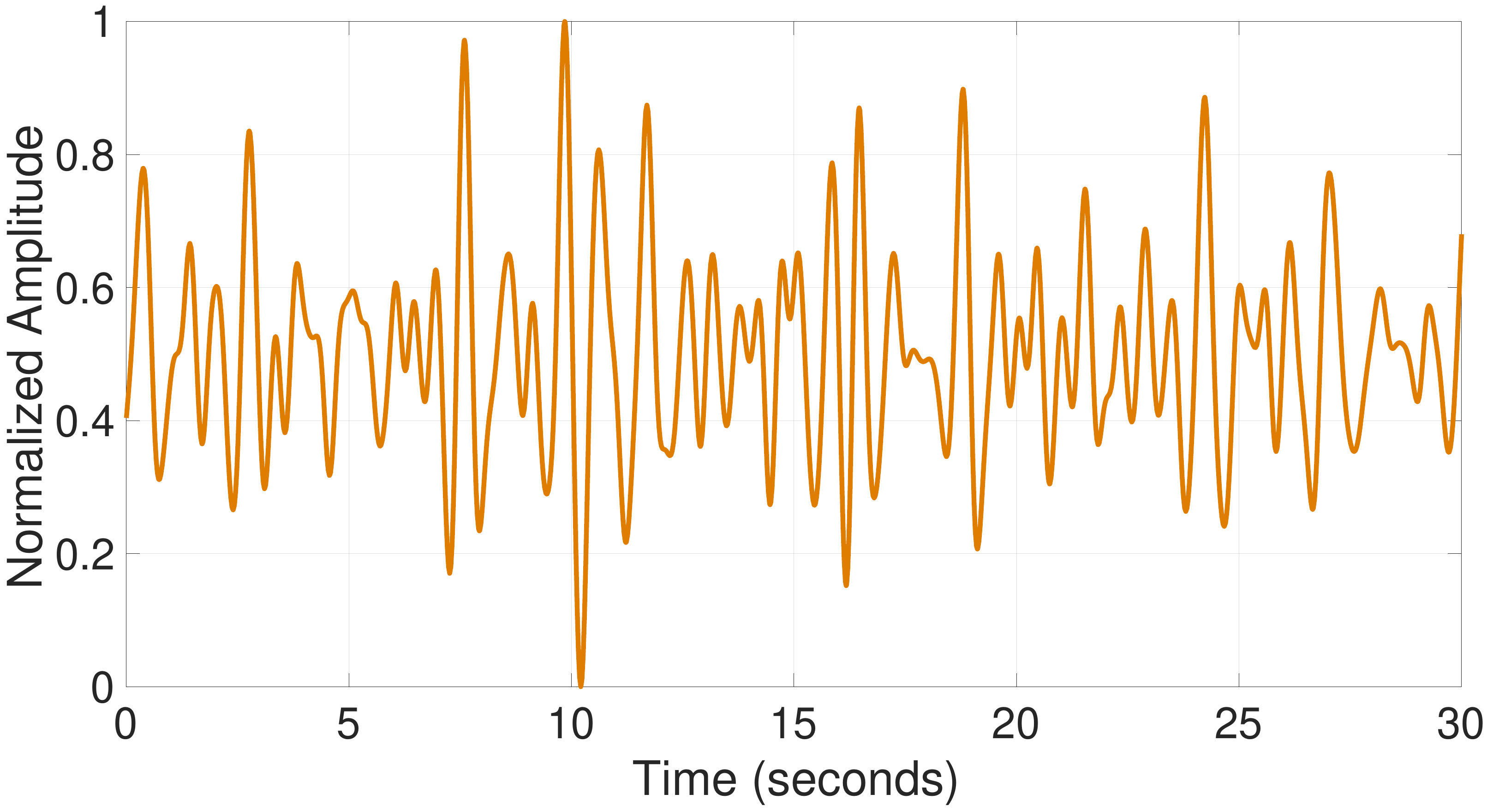}
		\subcaption{Extracted heartbeat waveform}
	\end{subfigure}
	\caption{\bb{Location-based vital sign monitoring based on 5 GHz WiFi signals.}}
	\label{Fig_vital_sign}
	\vspace{-1.5em}
\end{figure*}

\subsection{Human Sensing Case Study}
In the following, we present two case studies: random phase removal and vital sign estimation, to illustrate the practical sensing capabilities enabled by CSI signal processing.

\subsubsection{Random Phase Removal} 
We conduct an experiment to demonstrate the effect of random phase removal using different methods. An NI Massive MIMO testbed serves as the base station (BS), and a USRP device acts as the user equipment (UE). \bb{The BS and UE are deployed in a typical indoor laboratory environment containing static multipath components from walls, floors, desks, chairs, and monitors. The transmitter and receiver are placed 2.2 m apart, with the transmitter positioned at an angle of $50^{\circ}$ relative to the receiver antenna array.} Uplink CSI is collected from 3.1 GHz LTE pilot signals with 100 subcarriers over a 20 MHz bandwidth. During the experiment, \bb{a human target moves along a 2-meter by 3-meter rectangular trajectory, with the distance to the transmitter varying between 2 m and 5 m}. \bb{Fig.~\ref{Fig_phase_removal} shows Doppler–time heatmaps illustrating the impact of random phase removal, generated using raw CACC, CASR, CACC-variant, and single-antenna methods. Since the Doppler shifts are similar across antennas, applying raw CACC, i.e., computing the conjugate product between CSI measurements from two antennas, produces mirror-like Doppler components at both $+f^D$ and $-f^D$ with comparable amplitudes, as can be seen from Fig.~\ref{Fig_phase_removal}(a). This leads to ambiguity in distinguishing whether the target is approaching or moving away, as reflected by the symmetric Doppler signatures along the vertical axis. In contrast, the multi-antenna CASR and CACC-variant methods in Fig.~\ref{Fig_phase_removal}(b) and Fig.~\ref{Fig_phase_removal}(c) effectively suppress these mirror components, enabling more accurate Doppler extraction. However, CASR disrupts the inherent linear relationships between delay and AoA in the CSI data, which complicates subsequent feature extraction steps. On the other hand, the CACC-variant approach requires at least three receiving antennas. For the single-antenna approach, only the delay and Doppler dimensions are available. In this case, a reference signal is constructed from the delay domain to estimate relative TO and CFO across CSI measurements. Then, their effects can be eliminated by computing the conjugate product between the reference signal and the original CSI. As a result, random phase distortions can be suppressed even in single-antenna configurations, shown in Fig.~\ref{Fig_phase_removal}(d). This method leverages both temporal and spectral domain features, but its performance is inherently constrained by the limited bandwidth.} The results demonstrate that the proposed methods effectively resolve Doppler ambiguity and yield clearer, more stable Doppler signatures. 

\subsubsection{Location-based Vital Sign Estimation} 
In this experiment, a human subject remains seated in a static indoor environment. A 5 GHz WiFi signal is utilised for CSI collection, with a 1TX-3RX configuration based on Intel 5300 NICs. The system operates at a centre frequency of 5.32 GHz, with a CSI sampling rate of 1 kHz. \bb{The transmitter and receiver are placed 2 m apart with a LOS path, and both are deployed in the same laboratory environment. The vertical distance between the subject and the transceiver is approximately 4 m.} As illustrated in Fig. 3, the proposed location-refined vital sign sensing approach enables the extraction of both respiration and heartbeat signals from the residual CSI phase. \bb{We first leverage the method described in Section III. C to estimate the target's position in the Doppler, delay, and AoA dimensions. Since chest movements caused by respiration induce Doppler shifts, we apply a sliding window of 2 s with a step size of 0.1 s. Within each window, we remove the CSI mean to suppress static clutter and apply FFT along the temporal axis to obtain Doppler representations. The target bin with the largest magnitude is selected as the dominant Doppler component. Based on this, additional FFTs are applied along the delay and AoA dimensions to further suppress interference. The phase of the dominant bin across all dimensions is then extracted, as shown in Fig.~\ref{Fig_vital_sign}(a). We then perform phase differencing and apply a low-pass filter with a cutoff frequency of 2 Hz to extract the respiration waveform, as shown in Fig.~\ref{Fig_vital_sign}(b). Some sharp peaks observed in the waveform are attributed to heartbeat-induced micro-vibrations. As shown in Fig.~\ref{Fig_vital_sign}(c), we apply a bandpass filter in the 0.8-2 Hz range and a smoothing process to isolate the heartbeat signal. The estimated respiration rate is approximately 22 breaths per minute, consistent with manual counting. The heart rate is around 75 beats per minute, matching the measurement from a smartwatch. These results confirm the feasibility of non-contact physiological monitoring using commodity WiFi hardware.}

\section{ISAC for Environmental Sensing}

ISAC has recently been extended beyond vehicular and industrial applications into environmental domains, enabling low-cost, large-scale monitoring using existing wireless infrastructure. By extracting environmental signatures embedded in ambient communication signals, ISAC supports scalable sensing to obtain rainfall, soil moisture, and water level data. In the following, we present key contributions in these domains and report a real-world water sensing case study.

\subsection{Overview of ISAC Environmental Sensing Works}
\subsubsection{Rainfall Sensing}  
Rainfall-induced signal attenuation provides a natural mechanism for opportunistic rainfall estimation using existing wireless infrastructure. ISAC studies in this area commonly rely on the empirical A–R relationship, $A = a R^b$, linking signal attenuation $A$ (in dB/km) to rainfall rate $R$ (in mm/h), with parameters calibrated to frequency and drop size distribution. Field campaigns using mmWave links (e.g., 25–38~GHz) have demonstrated the ability to differentiate between convective and stratiform rain types based on link dynamics \cite{han_characteristics_2021}.

Recent approaches combine time-series signal features (e.g., RSSI fade slope, link asymmetry) with deep learning models such as recurrent neural networks (RNNs) to enhance spatial and temporal resolution \cite{jacoby_integrated_2024}. Urban-scale link tomography has also been used to reconstruct rainfall fields using incomplete or lossy data streams \cite{zohidov_tomographic_nodate}. Some lightweight methods exploit LTE/5G signal metrics (e.g., RSRP, RSSI) from mobile devices to detect rainfall onset with high accuracy in edge-based setups \cite{sabu_effect_2017,beritelli_rainfall_2018}. While promising, current ISAC rainfall sensing remains limited by multipath interference, the narrow dynamic range of power-only metrics, and the need for frequent recalibration across regions.

\subsubsection{Soil Moisture Sensing}  
Soil moisture changes the dielectric permittivity of soil, which modulates RF propagation delay, phase, and reflectivity—making it a natural candidate for ISAC sensing. Theoretical models such as Dobson and Mironov relate volumetric water content $\theta$ to $\varepsilon(\theta)$, which in turn influences observable features like phase shift ($\Delta \phi \propto \sqrt{\varepsilon}$) and interface reflectivity ($\Gamma$). These relationships form the basis for reflectometric and delay-based sensing.

On the system side, LoRa-based soil probes using RSSI-phase correlation have achieved sub-4\% moisture error in testbeds \cite{chang_sensor-free_2022}, while passive CSI-based WiFi platforms such as SoilTAG extract moisture-sensitive subcarrier profiles for in-situ tracking \cite{jiao_detecting_2022}. Ambient LTE downlink signals have also been reused for passive sensing using matched filtering and RSRP monitoring \cite{feng_lte-based_2022}, offering power-efficient alternatives in agricultural deployments. At larger scales, GNSS reflectometry (GNSS-R) has shown strong sensitivity to both surface and root-zone soil moisture in diverse terrain, with RMSEs under 0.05~cm$^3$/cm$^3$ in some campaigns \cite{chew_soil_2018,el_hajj_ground-based_2024}.

Despite these advances, ISAC soil moisture sensing faces challenges related to soil type heterogeneity, and changing surface conditions. Addressing these issues will likely require fusing RF features from different systems, supported by physically informed learning models.

\subsubsection{Water Sensing}  
Compared to rainfall and soil moisture, ISAC-based water level sensing is less mature but increasingly relevant for flood detection and hydrological monitoring. Water surfaces affect signal reflection paths, Doppler spread, and phase coherence—especially in low grazing-angle or NLoS settings. Simulation studies and field experiments have shown that fluctuations in CSI magnitude, time-of-flight (ToF), and beam directionality can indicate changes in surface water height \cite{ren_liquid_2020,peters_simulations_2024}. 
To illustrate the practical feasibility of ISAC water sensing, we present a field study using passive LTE-based water sensing conducted at a riverside test site in Sydney, Australia. This is detailed next.

\subsection{LTE-Based Water Sensing Case Study}
\begin{figure}[!t]
	\centering
	\includegraphics[width=8cm]{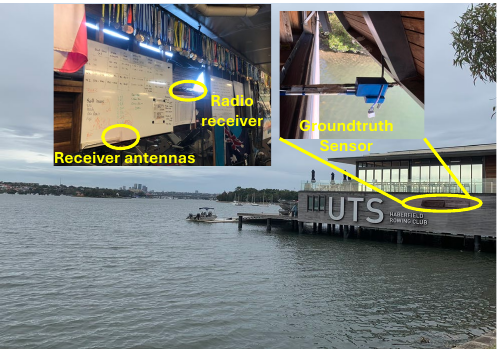}
	\caption{Experimental water sensing deployment at the UTS Haberfield Rowing Club. {A passive SDR receiver with two monopole antennas are placed indoors, and a sonar-based water level sensor is installed under the wharf structure. Downlink signals from multiple LTE base stations are captured by the SDR receiver in NLoS conditions.}}
	\label{fig:club_water_sensing}
\end{figure}

To investigate the feasibility of ISAC-style passive water sensing, a field campaign was conducted at the University of Technology Sydney (UTS) Haberfield Rowing Club (Fig.~\ref{fig:club_water_sensing}), located alongside the Parramatta River in Sydney, Australia. The setup comprised a software-defined radio (SDR) receiver and a high-precision sonar-based water level sensor. 
{As shown in Fig. \ref{fig:club_water_sensing}, simple monopole antennas, as attached to an indoor whiteboard, are used for the SDR receiver.} 
Operating in a NLoS geometry relative to surrounding infrastructure, the SDR nonetheless captured strong downlink signals from seven LTE cells, spanning four major bands commonly used in Australia (Bands 1, 3, 5, and 28).
This passive, infrastructure-agnostic deployment enabled rich multi-band, multi-cellular signal observation over natural tidal cycles, offering a realistic testbed for assessing ambient RF sensing capabilities in dynamic riverine environments.

\begin{figure}[!t]
	\centering
	\includegraphics[width=8cm]{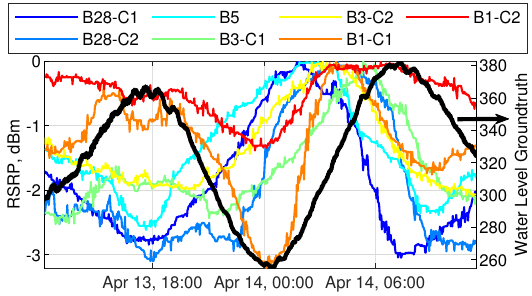}
	\caption{RSRP traces from LTE carriers overlaid with sonar ground-truth water level. Variations in correlation strength across bands reflect different propagation sensitivities to tidal dynamics. Bn denotes Band-n, with n taking 1, 3, 5 and 28, and Cm denotes Cell-m, where m is the index of different cells identified for each band.}
	\label{fig:rsrp_water}
\end{figure}

Conversely, signals from higher-frequency bands (e.g., Band 1 and Band 3) demonstrate weaker or lagged responses, possibly due to their limited interaction with water-reflected or refracted paths. These differences illustrate that environmental sensitivity varies across frequency bands and cell geometries. Exploiting such multi-band diversity can therefore improve the robustness and adaptability of ISAC-based water sensing systems.

\begin{figure}[!t]
	\centering
	\includegraphics[width=8cm]{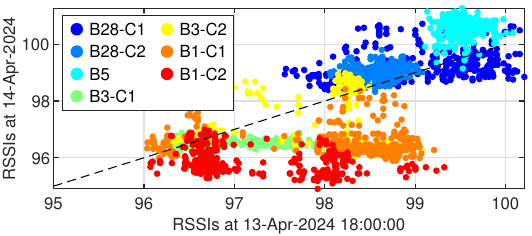}
	\caption{Scatterplot of LTE RSSIs (1~ms subframe samples) measured at 18:00 (13-Apr) and 00:00 (14-Apr). Displacements from the diagonal capture significant propagation changes linked to water level variation.}
	\label{fig:rssi_comparison}
     \vspace{-1.5em}
\end{figure}

To investigate short-term environmental influences, we analysed subframe-level RSSI samples captured at two distinct times separated by six hours. Fig.~\ref{fig:rssi_comparison} plots the measured RSSIs for seven PCIDs across the two snapshots. While perfectly stable propagation would align points along the diagonal, the observed spread and centroid displacement clearly indicate time-varying signal characteristics.
Given that these temporal shifts coincide with the changing tide (confirmed from sonar measurements in Fig. \ref{fig:rsrp_water}), the results suggest that even short-term, small-scale water level variations can perturb multipath structures sufficiently to be captured by LTE signals. This supports the hypothesis that water-induced changes in propagation delay, scattering, and surface reflection are detectable even under NLoS conditions.

\begin{figure}[!t]
	\centering
	\includegraphics[width=8cm]{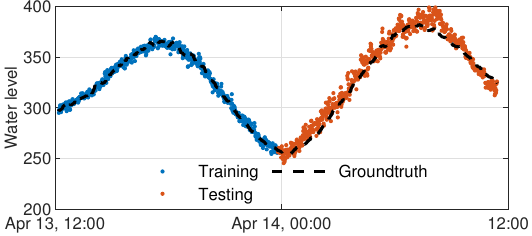}
	\caption{Estimated vs. ground-truth water level using a linear regression model trained on LTE RSRP signals and meteorological features. A RMSE of 7.36~cm was achieved during testing.}
	\label{fig:lte_results}
     \vspace{-1em}
\end{figure}

Building on these feature observations, we initially developed a regression-based model to predict water level using low-level LTE metrics. A simple linear regression was trained on 12 hours of feature data—including RSRP measurements from seven LTE carriers to predict the water level at 10-minute intervals in the next 12 hours.
As shown in Fig.~\ref{fig:lte_results}, the model output closely tracks the sonar ground truth during the 12-hour testing phase, achieving a root-mean-square error (RMSE) of 7.36 cm. The relatively low error and consistent tracking despite the use of basic linear models suggest that ambient RF signals carry sufficiently strong environmental signatures for accurate sensing, even without complex preprocessing or extensive feature engineering.

\begin{figure}[!t]
	\centering
	\includegraphics[width=8.5cm]{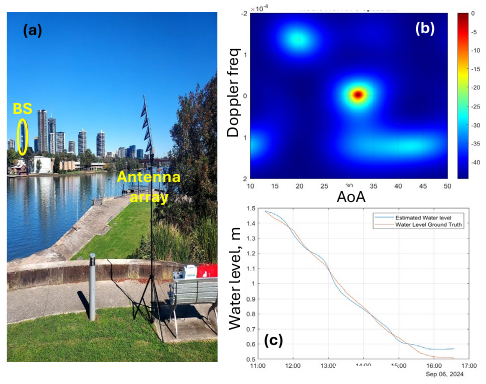}
	\caption{Illustration of the current water sensing research setup and results \cite{fahad_iotj_water}. 
		(a) Field experiments at the Paramatta River, Sydney, Australia, showing the antenna array receiving downlink signals from a remote base station (BS) near 300m away. The antenna is about 100m away from the river bank; 
		(b) Joint Doppler-AoA spectrum obtained via advanced space-time processing, revealing signal components associated with water and other scatterers; 
		(c) Estimated water level over time compared with ground truth measurements, demonstrating high accuracy with sensing resolution on the order of $10^{-4}$~m$^2$.}
	\label{fig:current_research}
	\vspace{-1em}
\end{figure}

{Building on the promising results of initial experiments, our recent research has expanded into several directions aimed at improving the sensitivity, robustness, and generalisability of ISAC water-level sensing. As illustrated in Fig.~\ref{fig:current_research}, a prototype system was tested at Rhodes Waterside, along the Parramatta River in Sydney, Australia, where downlink cellular signals from a distant base station were captured by a ground-based antenna array. Through advanced space-time signal processing, the system extracted joint Doppler–AoA features (Fig.~\ref{fig:current_research}(b)) and achieved highly accurate water-level estimates that closely match ground truth measurements (Fig.~\ref{fig:current_research}(c)); refer to \cite{fahad_iotj_water} for more technical details.

To further enhance performance, finer-grained physical-layer features, such as CSI amplitude and phase profiles, Doppler spread, and polarisation, are being explored, with some interesting results presented in \cite{wang2025water}. In addition, fusion of heterogeneous signal sources, including LTE and WiFi, is under active development to improve spatial and temporal sensing resolution. Complementing these efforts, physics-informed machine learning models are being integrated to support robust generalisation across varying environmental conditions, channel dynamics, and infrastructure layouts.
Together, these effort represent an active and evolving research paradigm toward scalable, infrastructure-leveraged ISAC solutions for environmental sensing. 
}

\section{Concluding Remarks and Open Challenges}

This paper has presented the first unified analysis of ISAC-enabled sensing across both human and environmental domains. By bridging near-field, human-centric applications such as motion tracking and vital sign monitoring with far-field environmental sensing tasks including rainfall estimation, soil moisture detection, and water level monitoring, we highlighted the common physical mechanisms through which diverse phenomena affect wireless signal propagation. These mechanisms give rise to measurable signatures in amplitude, phase, Doppler, and channel statistics that can be exploited through both physics-driven models and learning-based inference.

We reviewed representative works across both domains, identifying the distinct propagation conditions, sensing features, and system requirements that characterise each. Through field experiments involving passive LTE signal capture, we demonstrated the feasibility of using ambient RF infrastructure for real-time water sensing in non-line-of-sight scenarios. These findings validate the practical relevance of ISAC sensing beyond theoretical or simulation settings.

\bb{A key insight from this cross-domain review is that, despite task-specific implementations, many ISAC sensing systems operate on shared physical mechanisms, essentially the signal changes caused by environmental or physiological dynamics. This suggests that ISAC need not remain a fragmented collection of siloed applications. Instead, we argue that ISAC is poised to evolve into an integrated sensing layer embedded within communication infrastructure, which is capable of supporting diverse tasks through modular signal acquisition, unified feature representations, and adaptable inference strategies. While existing deployments often target isolated use cases, our unified framework illustrates how a general-purpose ISAC system can emerge from common building blocks across domains. Eventually, sensing applications can be realised as APPs running on base station or phone sides. This, however, will require progress in abstraction, coordination, and context-aware system design.}

Looking forward, several open challenges are identified, as detailed below. 

\bb{
\subsubsection{Human Sensing under Weak LOS and NLOS Scenarios} 
Most existing ISAC systems rely on a static LOS path between the transmitter and receiver to extract key features such as Doppler, delay, and AoA for passive human sensing. However, this assumption often breaks down in practical indoor environments, where the LOS path may be blocked or completely absent. For example, sensing indoors using mobile signals generally does not have a LOS path, and WiFi sensing indoors with transceivers in different rooms can suffer from complex multiplex propagation environments.
In such weak LOS and NLOS scenarios, effectively suppressing interferences poses a significant challenge. 
}

\bb{\subsubsection{Multi-Target Human Sensing Scenarios}
Accurate extraction and tracking of multiple human targets remain a significant challenge, particularly in low-bandwidth communication systems where the delay resolution is inherently coarse. Additionally, certain random phase removal methods may distort useful signal components, further complicating multi-target separation. Compared to the delay domain, Doppler and AoA features can provide effective discrimination between multiple moving targets. These two domains enable the detection of potential targets even when delay resolution is insufficient. 
By integrating Doppler, AoA, and delay information into an EKF-based framework \cite{wang2024passive}, it becomes feasible to achieve robust multi-target tracking.}

\bb{
\subsubsection{Robustness across Diverse ISAC Deployments}
ISAC systems vary significantly in terms of centre frequency, bandwidth, antenna configurations, and deployment geometries, making it challenging to generalise models across different setups. Models trained under one configuration often fail to perform well when applied to other domains due to hardware-dependent discrepancies and spatial-temporal mismatches. To address this, a unified 3D feature representation based on Doppler–Delay–AoA signatures can be employed. This representation abstracts away hardware-specific variations and captures motion-induced changes in a format that is more universal and transferable. 
Such abstraction enables shared model training, domain adaptation, and large-scale data augmentation, thereby facilitating the development of more robust ISAC sensing models across diverse system setups.}

\subsubsection{Antenna and Platform Variability}
Directional or embedded ISAC systems (e.g., on UAVs, vehicles, or wearables) require reconfigurable antennas. Rainfall sensing via CMLs has shown that tilt, antenna tilt and wetting, and material absorption introduce systematic bias \cite{zinevich_prediction_2010,van_leth_urban_2017}. Similarly, for soil moisture, antenna–ground coupling is highly dependent on soil type and depth, if the antennas are close to the ground \cite{skiljo_self-sensing_2023}. Future designs should prioritise beam agility, polarimetric calibration, and orientation-agnostic response, potentially leveraging metamaterial or self-sensing antenna technologies. Emerging solutions such as generalised joint coupler antennas \cite{guo2022optimization,guo2023multibeam,wu2023deterministic} offer greater configurability than traditional multibeam architectures, supporting sensing and communication needs dynamically. Leaky-wave antennas \cite{chen2019polarization,chen2023wideband} further offer promising frequency-dependent beam steering capabilities, reducing the need for multiple RF chains and simplifying hardware complexity in multiband ISAC deployments. 

\bb{In addition, low-cost, lightweight, and energy-efficient antenna technologies will be essential for enabling scalable and sustainable ISAC deployments, particularly in mobile and resource-constrained platforms. Innovations such as printed antennas, textile-based antennas, and origami-inspired foldable arrays offer promising paths toward compact, conformal, and low-power solutions. Likewise, green antenna technologies, including recyclable materials, passive or self-powered designs, and energy-harvesting capabilities, align with broader goals of environmental sustainability and reduced lifecycle carbon footprint.
}

\subsubsection{Task-Aligned Feature Engineering and Multi-Modal Fusion}

Effective environmental sensing requires extracting domain-relevant features—such as amplitude attenuation for rainfall, phase delay for soil moisture, and backscatter variations for water levels—often buried within infrastructure-specific CSI or RSSI logs. To generalise across settings, interpretable metrics like Doppler variance, path coherence, and delay spread entropy must be derived. Physics-informed learning models, such as I-RNNs with A–R constraints \cite{jacoby_integrated_2024}, help bridge raw signal features and environmental parameters more robustly than black-box approaches. In parallel, multi-modal fusion, leveraging cellular signals, weather radar, GNSS-R, satellite IR, and IoT sensors, can enhance spatial resolution and robustness. Studies have shown that combining radar with commercial microwave links or satellite data significantly improves rainfall estimation accuracy \cite{lombardi_combined_2024, tichavsky_multimodality_2017}. However, the fusion of asynchronous and resolution-mismatched data remains a challenge. Recent advances in graph neural networks, attention-based transformers, and cross-domain latent embeddings offer promising pathways, particularly under semi-supervised or weakly labelled conditions \cite{lu_deep-learning-based_2024}.

\subsubsection{Data Scarcity and Generalisation}

Environmental ISAC systems also face two key challenges: data scarcity and limited generalisability. In rural or under-instrumented regions, labelled data is often lacking, while even in urban settings, variations in weather, soil, and infrastructure demand region-specific model adaptation \cite{bonkoungou_determination_2024, ostrometzky_opportunistic_2024}. To address this, approaches such as synthetic data generation via ray-tracing or rainfall emulators, perturbation of measured signals, and unsupervised domain adaptation are being explored. Federated learning further enables collaborative, privacy-preserving model training across distributed nodes \cite{lu_deep-learning-based_2024}. Complementarily, recent advances in wireless knowledge representation suggest that large language models (LLMs) can transform signal-derived features into high-level semantic insights, such as flood warnings, irrigation needs, or infrastructure risks \cite{zhang_wireless_2024}. This vision of hierarchical signal interpretation requires bridging low-level quantitative measurements with symbolic reasoning, a challenge that early work in vision–language and hybrid embeddings is beginning to tackle. Addressing these challenges is critical to moving ISAC from isolated proofs-of-concept to scalable, intelligent systems for pervasive environmental awareness in 6G and beyond.

\ifCLASSOPTIONcaptionsoff
\newpage
\fi
	
\bibliographystyle{IEEEtran}
\bibliography{IEEEabrv,./bib1/rainfall, ./bib1/env_sensing_overall,./bib1/phone,./bib1/soil, ./bib1/water, ./bib1/antenna, ./bib1/isac_background, ./bib1/refs_zq}

\end{document}